\documentclass[aps,prl,twocolumn,floatfix]{revtex4}

\usepackage{amssymb}
\usepackage{amsbsy}
\usepackage{amsmath}
\usepackage[pdftex]{graphics}
\usepackage{graphics}
\usepackage{graphicx}
\usepackage{dcolumn}
\usepackage{subfigure}
\usepackage{bm}
\def\be{\begin{equation}}
\def\ee{\end{equation}}

\begin{document}

\title{Switching the anomalous DC response of an AC-driven quantum many-body system}

\author{Arnab Das and R. Moessner}

\affiliation{Max-Planck-Institut f\"{ur} Physik komplexer Systeme,
N\"{o}thnitzer Strasse 38, Dresden, Germany
}
\begin{abstract}
For a class of integrable quantum many-body systems, symmetric AC 
driving can generically produce a steady DC response. We show how such 
dynamical freezing can be switched off, not by forcing the 
system to follow the (arbitrarily fast) driving field, but rather through 
a much slower but complete oscillation of each individual mode of the 
system at a frequency of its own, with the slowest mode exhibiting a 
divergent period. This switching can be controlled in 
detail, its sharpness depending on a particular parameter of the 
Hamiltonian. The phenomenon has a robust manifestation 
even  in the few-body limit, perhaps the most promising setting for 
realisation  within existing frameworks.
\end{abstract}

\maketitle


\noindent


The coherent non-equilibrium dynamics of driven quantum systems offers a rich and largely unchartered field of many-body physics. 
While the physics of a periodically driven single quantum particle
has remained a lively topic of interesting research through past decades  
(see, e.g., \cite{Dunlap,Hanggi-1,Hanggi-2,Andre-2,Andre-3}), recent studies in periodically driven 
quantum many-body system have revealed surprising phenomena like
dynamical many-body freezing \cite{AD-DQH,AD-BDD,Kris-Bose-Hubbard}, appearance of slow solitary 
oscillations \cite{AD-BDD}, superfluid-insulator and localization transition due to effective renormalization 
of the hopping \cite{Andre-1, Andre-1a, Gil} which often drastically fail to conform to our classical intuition.
Here we add a novel phenomenon to this collection, that of regime switching: at its most extreme, simply tuning a 
parameter in the Hamiltonian can lead to a transition between a completely frozen and an entirely dynamic regime.

Our study contributes to the broader question of how quantum systems respond to non-adiabatic driving. 
A conceptually simple setting is provided by a quantum system driven periodically in time with a purely 
AC driving of frequency $\omega$, with $m_z(t)$ denoting the ``response'' -- a local quantity that varies 
in time under the influence of the driving. Let us assume \cite{Non-Symmetric} that the equilibrium value of this quantity
vanishes when the driving field is zero, and that, like the driving, it is symmetric about zero over a driving period. 
Starting from the equilibrium state, purely AC driving must produce zero DC response over each cycle in the 
adiabatic limit (provided it exists, i.e. that there is no level crossing). However, if the driving is non-adiabatic, 
the response can exhibit an average non-zero DC contribution even in the limit of infinite driving time. 
In an extreme case, quantum interference may dynamically stabilize the response around a nonzero DC value \cite{AD-DQH,AD-BDD,Bastidas}. 
The (non-adiabatic) anomalous DC response is then defined as    
\be 
Q = \lim_{\tau\rightarrow\infty}\frac{1}{\tau}\int_{0}^{\tau}m_{z}(t)dt  \ .
\label{Q-defn}
\ee 
\noindent  
$Q$ can have strong 
non-monotonic behavior with a peak-valley profile in the space of driving parameters,
frequency $\omega$ and amplitude $h_{0}$ \cite{AD-DQH,AD-BDD}. 
For certain combinations of $h_{0}$ and $\omega$, the response  
remains frozen extremely close to its initial value for all time for any initial state. 

Here we discuss a very different but complementary phenomenon -- how a non-zero $Q$ can 
be made to disappear by simply tuning a coefficient in the Hamiltonian. This switching between $Q$ finite and 
$Q = 0$ is associated with complete transfer of population between quasi-particle levels due to the driving,
 which surprisingly happens on very long timescales even when the driving is arbitrarily fast.
Under the special condition of maximal freezing, the transition approaches a discontinuous limit. 
The phenomenon has a dramatic precusor already in the limit of
small system-sizes ($\agt$ 6 spins). In fact, conditions for switching are less restrictive in this
limit, and hence more easily observable in experiment. 

In the remainder of this paper, we first summarise the key observations for a quantum spin chain in a 
time-varying field. We then provide a detailed analytical theory, which agrees with a numerically exact 
treatment very well in the limit of fast driving. This is followed by an account of a finite-size version 
of the switching effect, together with a discussion of experimental realisability, and a concluding outlook.\\

\noindent
{\it {\bf Regime switching:}} 
\begin{figure*}
\centerline{
\includegraphics[width=0.32\linewidth, angle=0]{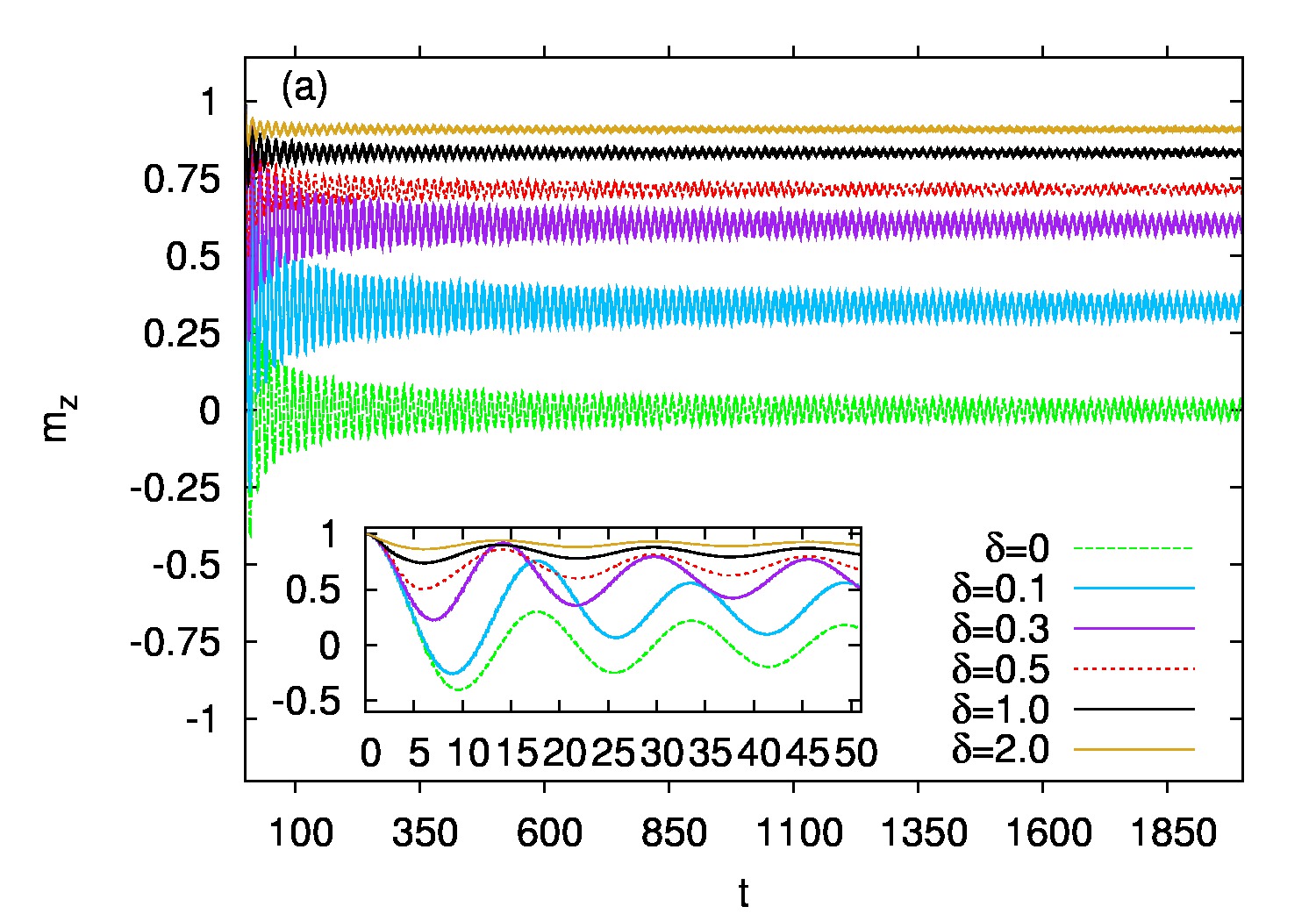}
\includegraphics[width=0.35\linewidth,angle=0]{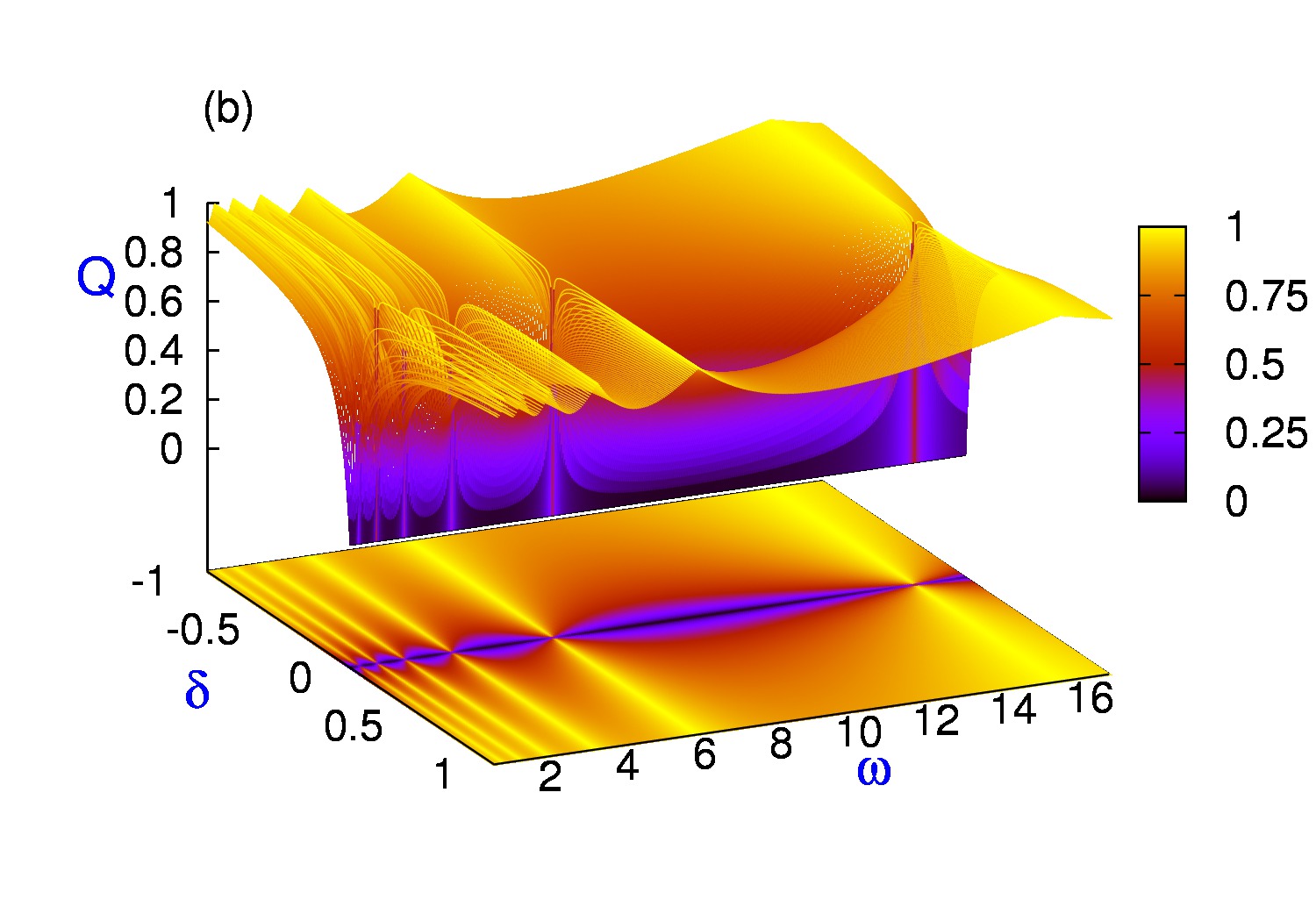}
\includegraphics[width=0.32\linewidth, angle=0]{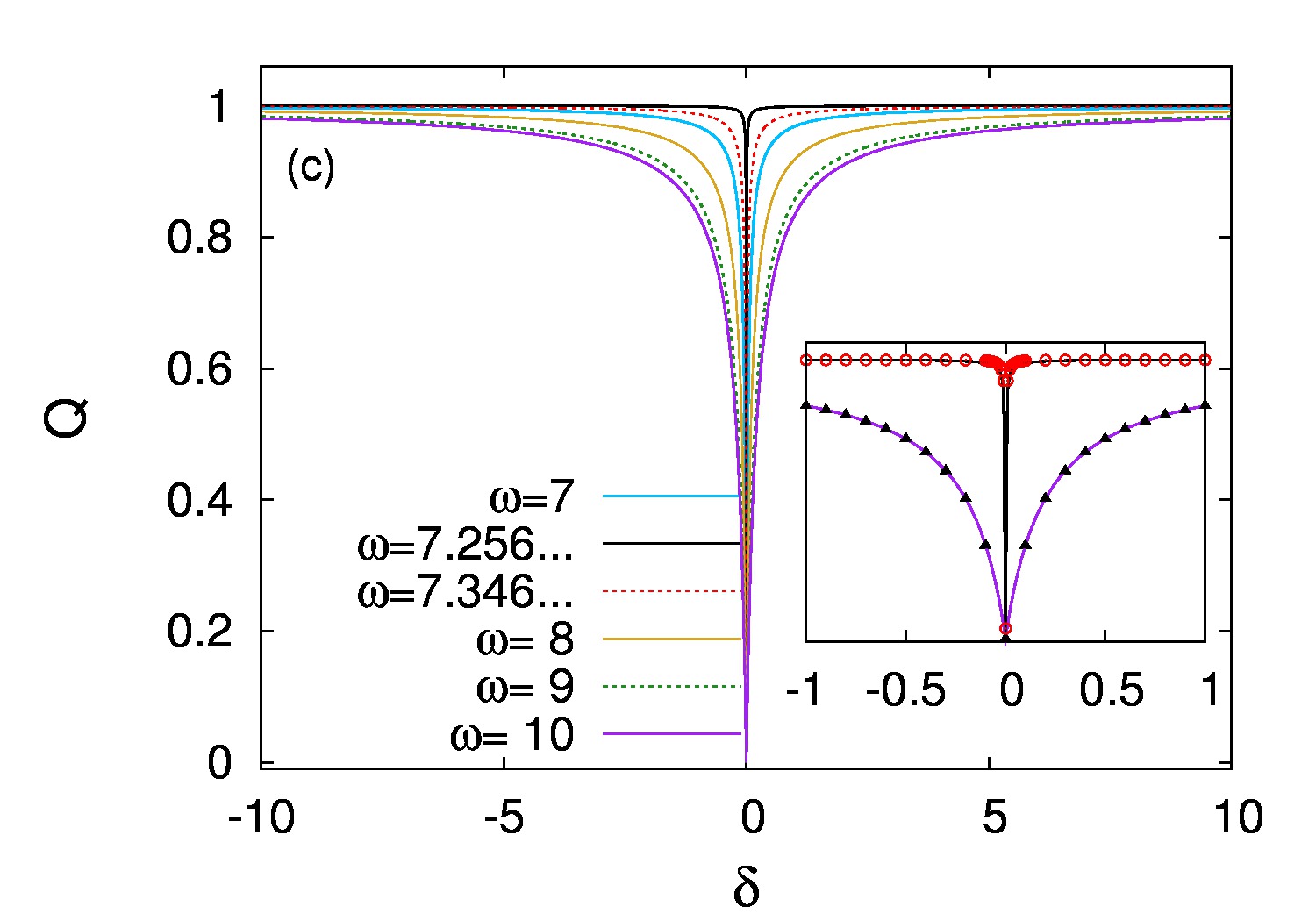}
}
\caption{\small{\bf{Dynamical transition:}} 
{\bf (a):} Exact numerical result for $m_z$ vs $t$ for the quantum XY chain with $\omega = 10$ over several periods
for $L = 10^{4}$. Inset shows the initial relaxation. We have taken $h_{0} = 20$ and $\gamma = 0.5$ 
(throughout the paper, unless otherwise noted). 
{\bf (b):} 3d profile of $Q$ vs $\delta$ and $\omega$. The sharp drop in $Q$ at around $\delta = 0$
(projected to the violet line on the base) and the peaks along the $\omega$ line (projected to bright yellow
stripes on the base) are visible.
{\bf (c):} $Q$ vs $\delta$  for different $\omega$ for $h_{0} = 20$. The figure shows sharp drop of 
$Q$ at $\delta = 0$. The sharpness is a non-monotonic function of $\omega$, approaching a discontinuous 
limit for certain values of $\omega$ identified as the peak frequencies. The inset shows comparison between 
analytical formula (Eq. \ref{Q-ana}) for an infinite system with numerics for $L = 100$, averaged over $> 10^4$ cycles for
$\omega = 10$ (broader) and $\omega = \omega^{P_{k}}_{0} + 10^{-2} = 7.256...$ (sharper).  
}
\label{Q-vs-delta}
\end{figure*}
Consider the Hamiltonian
\begin{equation}
H_{XY} =
-\frac{1}{2}\left[J_{x}\sum_{i=1}^{L} \sigma_{i}^{x}\sigma_{i+1}^{x} 
+ J_{y}\sum_{i=1}^{L} \sigma_{i}^{y}\sigma_{i+1}^{y}
+ h_{z}(t)\sum_{i=1}^{L} \sigma_{i}^{z}\right] 
\label{HXY}
\end{equation} 
\noindent
with $\sigma^\alpha$ Pauli matrices, $h_{z}(t) =  h_{0}\cos{(\omega t)}$,
and
\begin{equation}
\gamma = J_{x} - J_{y} \quad {\rm and} \quad \delta = J_{x} + J_{y}~~.
\label{gamma-delta}
\end{equation} 
We follow the transverse magnetization $m_{z}(t) = \frac{1}{L}\sum_{i}^{L}\sigma_{i}^{z}$, taking
the initial state to be fully polarized in the $+z$-direction ($m_{z} = 1$). Driving is strong and rapid,
$\omega,h_{0} > |\gamma|,|\delta|$. 

Fig. \ref{Q-vs-delta} (a) shows $m_z (t)$ for a system of $L = 10^4$ spins from a numerical solution of the time-dependent Schr\"odinger equation. After a transient, $m_{z}$ oscillates about a non-zero
steady average, $Q$. This approaches zero as $\delta \rightarrow 0$. There is a concomitant substantial 
enhancement of dynamics over long time-scales, visible in the increasing amplitude of oscillations about $Q$.

\noindent
The dependence of $Q$ on driving parameters is shown in Fig. \ref{Q-vs-delta}(b),(c). Besides the modulation of $Q (\omega)$ at fixed $\delta$ \cite{AD-DQH,AD-BDD,Bastidas}, there is an even more pronounced dependence on $\delta$, with the response switched off, $Q (\omega, \delta = 0)\equiv0$ for vanishing $\delta$. The behaviour for small $\delta$ is rich, exhibiting a tunable slope, the value of which diverges at the frequencies of maximal freezing $\omega^P$, yielding a discontinuity there: 
$Q=0$ at $\delta=0$, and $Q\rightarrow 1$ as $\delta\rightarrow 0^{\pm}$.
   
Non-adiabatic evolution is of course not unexpected, since the gap closes at $h_{z}=\pm\gamma$ in every cycle; moreover, $\omega,h_{0}$ are much greater than the natural energy scales
$J_{x},J_{y}$ of the undriven system. However, disappearance of $Q$ and enhancement of
long-time dynamics as $\delta \rightarrow 0$  is surprising, particularly given the fact that reducing $\delta$ reduces the minimum gap, the smallest gap given by $|h_{0}\cos{\omega t} + \delta|$.


\noindent
{\it{\bf Analytical theory:}} 
We reduce Hamiltonian in (\ref{HXY}) to the direct sum of $2\times 2$ 
Hamiltonians in momentum space using Jordan-Wigner and Fourier transformations \cite{LSM}:
\be
H = \sum_{k}\psi_{k}^{\dagger}H_{k}(t)\psi_{k}; \ \  H_{k} 
= [h(t) + \delta f_{k}]\sigma^{z} + \Delta_{k}\sigma^{+} 
+ \Delta_{k}^{\ast}\sigma^{-} 
\label{Hk}
\ee
\noindent where $k$ is the quasi-momentum,
$\psi_{k} = (c_{1}^{\dagger}(k),c_{2}(k))$ is the spinor representing the fermionic
operators $c_{1,2}(k)$ in $k$-space, 
$\sigma^{\pm} = (\sigma^{x}\pm i\sigma^{y})/2$, 
 $\delta$ and $h(t)$ (independent of $k$) are the parameters in the many-body Hamiltonian,
and $\Delta_{k}= \gamma\sin{k}$ and $\delta f_{k}=\delta \cos{k}$ are functions of $k$. 
In the Schr\"{o}dinger picture $H_{k}$ represents the Hamiltonian for
the time-dependent wave function, which in turn is a direct product of
$k$-space wave functions: 
\be
|\psi(t)\rangle = \prod_{k}|\psi_{k}(t)\rangle = 
\prod_{k}[u_{k}(t)|0_{k}\rangle + v_{k}(t)|1_{k}\rangle].
\label{psi} 
\ee
\noindent
were $\{|0_{k}\rangle,|1_{k}\rangle\}$ are the basis states in which
the matrix form of $H_{k}$ is expressed in (\ref{Hk}). 

The Hamiltonian
in (\ref{Hk}) represents a class of systems including 
well known integrable 1D quantum models, such as the 
transverse Ising chain, 1D p-wave superconductor 
\cite{Kitaev-pwv, Diptiman-pwv}, or quantum spin ladders
\cite{Diptiman-ladder}. For instance, the Hamiltonian  
of a 1D p-wave superconducting wire of spin-less fermions
can be described by mapping $\delta$ onto the hopping, $\gamma$ onto the pairing gap, and $h(t)$ onto the chemical potential, with $m_z = 2n_f - 1$ related \cite{Kitaev-pwv}
\noindent
the average fermionic occupation number, $n_{f}$. A similar mapping is
applicable to the Kitaev spin ladder in \cite{Diptiman-ladder}.
%
%
%
%

The momentum space wave-function (\ref{psi}) satisfies the following time-dependent Schr\"{o}dinger equation
\be
i\frac{d|\psi(t)\rangle_{k}}{dt} = H_{k}|\psi(t)\rangle_{k}
\label{TDSE}
\ee  
\noindent
and the time-dependent transverse magnetization reads:
\be
m_{z}(t) = \langle\psi(t)|\sigma_{i}^{z}|\psi(t)\rangle
=  \frac{4}{N}\sum_{k>0} \left( |v_{k}(t)|^{2} - 1/2 \right)
\label{mz}
\ee

An exact analytical solution of Eq. \ref{TDSE} is not available, but for large $\omega$,
analytic solution is possible in the rotating wave approximation (RWA) \cite{AD-DQH, Nori}. 
Starting from a fully polarized ($m_z = 1$) initial state, corresponding to $v_{k}(t=0) = 1$ 
(a good approximation to the ground state of the Hamiltonian at $t = 0$
for $h_{0} \gg \gamma,\delta$),
\be
|u_{k}(t)|^2 =
1 - |v_{k}(t)|^{2} =
\frac{J_{0}^{2}(2h_{0}/\omega)\Delta_{k}^{2}}{\phi_{k}^2}\sin^{2}{(\phi_{k}t)}.
\label{Vk2}
\ee
\noindent where 
\be
\phi_{k} = \sqrt{J_{0}^{2}(2h_{0}/\omega)\Delta_{k}^{2}
+ \delta^{2}\cos^{2}{k}}~~.
\label{phik}
\ee
\noindent 
For periodic boundary conditions, we have wave vectors of the form $k = (2n+1)\pi/L$,
where $n$ denotes positive integers. Averaging over time and integrating over $k$ in the continuum ($L \rightarrow \infty$) limit, 
\be
Q = \frac{|\delta|}{|\gamma J_{0}(2h_{0}/\omega)| + |\delta|}~~.
\label{Q-ana}
\ee  
\noindent This formula (Fig.~\ref{Q-vs-delta}b) is one of our central results. 
It very accurately reproduces the numerical results  (Fig.~\ref{Q-vs-delta}c), 
in particular the switching off of $Q$ for $\delta=0$ unless the Bessel function simultaneously vanishes. 


%

\noindent
{\bf (Dis)appearance of $Q$:}
\noindent
From Eq. (\ref{Vk2}), the maximum amplitude of excitation for any given
mode is $\frac{J_{0}^{2}(2h_{0}/\omega)\Delta_{k}^{2}}{\phi_{k}^2}$, which is atmost unity,
and in general smaller. 
This is what leads to a finite $Q$ in our case,
as is apparent from the expression of $m_{z}$ in Eq. (\ref{mz}). Note here the role of the apparent 
``DC" part $\delta\cos{k}$ in the $2\times 2$ Hamiltonian (Eq. \ref{TDSE}), which plays the 
same role as would have been played by an additional DC part in the transverse field (albeit without the $k$-dependence). 
It is essential for the anomalous dc response: for $\delta = 0$, we have $Q = 0$ for all modes
and for all $\omega$, however large! 


It is important to note that $Q = 0$ emphatically does not reflect any adiabatic dynamics:
eqs.~(\ref{Vk2},\ref{phik}) encode the  drastic enhancement of dynamics of each 
$k$-mode at $\delta = 0$. At this point each mode has full oscillation with amplitude $|u_{k}|^2$ (but not frequency $\phi_k$)
 independent of both $k$ and parameters of the Hamiltonian. Thus,  the characteristic time of the full population transfer for a mode of quasi-momentum $k$,  
 $T_{k} = 2\pi/\phi_{k}$. Indeed, from Eq. (\ref{phik}) it follows that $T \to \infty$ for $\delta = 0$ and $\cos k \to 0$, so that some modes 
oscillate on diverging timescales, leading to an asymptotic 
$m_z (t) \sim J_0 \left( | 2 J_0 \left( \frac{2 h_0}{\omega} \right) | t \right) \sim \left| J_0 \left( \frac{2 k_0}{\omega} \right) t \right|^{-1}$. 
This form evidences the singular nature of the points $(\omega = \omega^P, \delta = 0)$, where $\omega^P$ are the zeroes of 
the Bessel function, as we discuss below.
 
\begin{figure}[h]
\centerline{
\includegraphics[width=0.65\linewidth, angle=0]{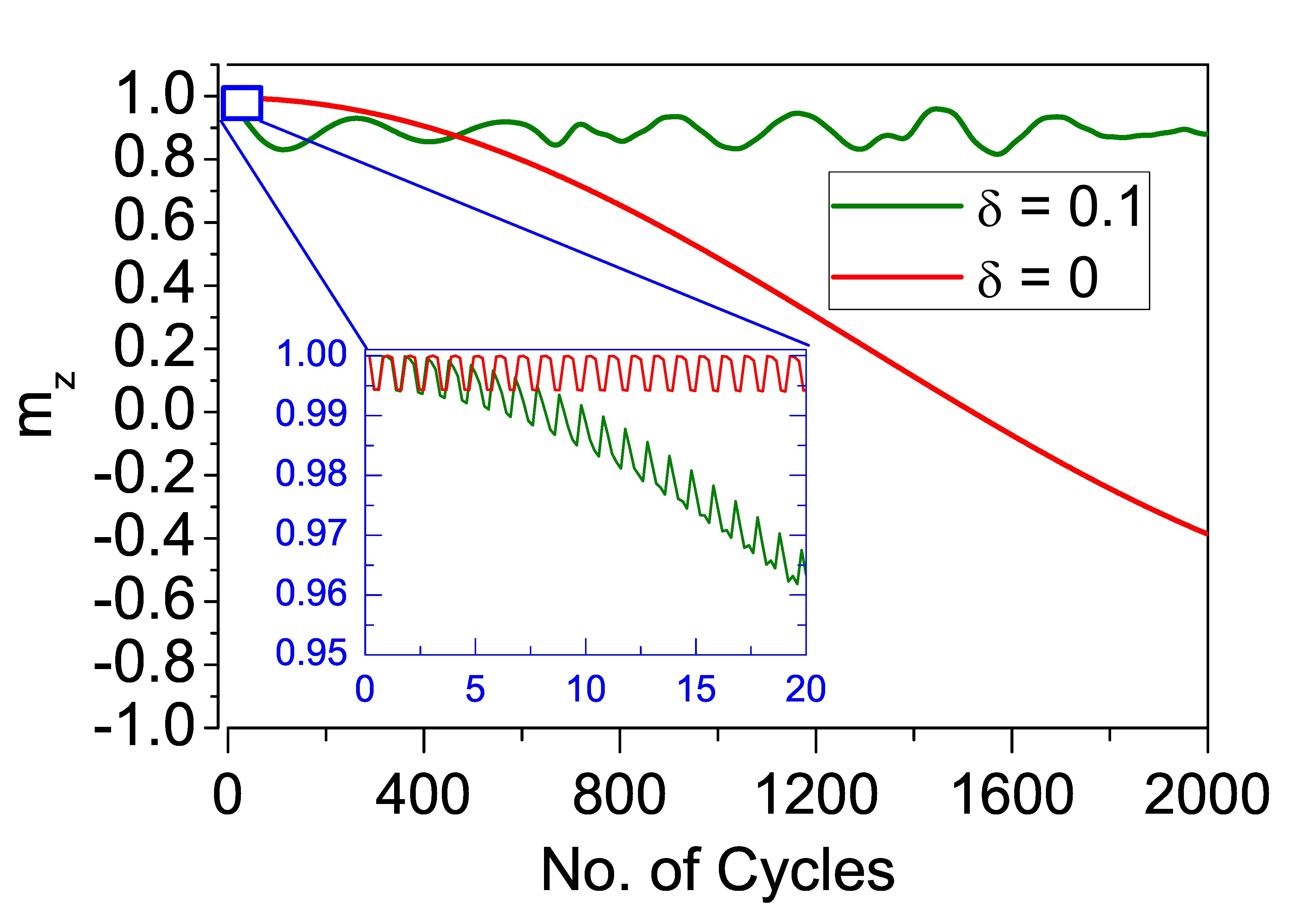}
}
\caption{\small{\bf{Short time vs long time dynamics:}}
Discrepancy between small short-time and large long-time excursions for the dynamical case, $\delta=0$, 
while, for $\delta = 0.1$, a fast initial decay is followed by long-time stability around the average $Q\neq0$ (numerical result for $L = 100$).
}
\label{Single-Sweep-Comparison}
\end{figure}

By contrast, the short-time behaviour offers little clues into what happens at long times (precluding a simple picture 
based on naive repetition of the single sweep result) and in fact makes the dynamical curves look more static than the frozen ones, 
in keeping with the observation that  switching involves times much longer than the sweeping period:  
Fig. \ref{Single-Sweep-Comparison} compares dynamics for $\delta = 0$ and $\delta = 0.1$ close to the sharp transition point
$\omega = \omega^{P}_{0} + 10^{-1}$. 
For $\delta = 0$, while the long-time dynamics takes the system
far away from the initial state, at the end of first few cycles, the system returns
remarkably close to the initial state. But for $\delta = 0.1$,    
the system remains quite close to the initial state for all later time, while its deviation 
from the initial state at the end of few initial cycles is systematically much larger than
that for the $\delta = 0$ case.

\noindent
{\bf Sharpness of the transition and the discontinuous limit:} 
\noindent 
As seen from Fig.\ref{Q-vs-delta} (b), (c), the sharpness of the transition varies non-monotonically
with $\omega$. For discrete values of $\omega$, the limit is
discontinuous: $Q=1$ for $\delta \ne 0$ and $Q=0$ for $\delta=0$.
As can be seen from Eq. (\ref{Vk2}), if $\delta \ne 0$, $Q = 1$ is maximal at the frequencies,
where $J_{0}(2h_{0}/\omega) = 0$, and, up to remarkable accuracy all the $k$-modes remain completely frozen for all time \cite{AD-DQH}. 
However, if $\delta = 0$, all modes undergo full oscillation.
Hence the limits $J_{0}(2h_{0}/\omega)\Delta_{k} \rightarrow 0$ and also $\delta\cos{k} \rightarrow 0$ do not commute: 
we either obtain $Q = 0$ or $Q > 1$, switching between complete dynamics and complete freezing.
The discontinuity is of course derived under RWA, whose accuracy increases with $\omega$, 
and it becomes exact as $\omega \rightarrow \infty$. 

The tunable sharpness of the switching might be interesting from the point of view of sensitive
detectors of very small static magnetic field. As discussed in the previous section, a small DC component of 
the transverse field plays the same role as a non-vanishing $\delta$. 
Now, if one sets $\delta = 0$, and tunes the system close to an $\omega_i^P$ then
a DC transverse field will result in a jump from $Q = 0$ to $Q = 1$.
Of course, the resolution time over which coherence needs to be maintained will grow as one approaches $\omega^P$. 
Since the transition persists in the limit of small system 
consisting of few spins (see below and the Supplementary Material), 
maintaining coherence over the period necessary for detecting even very weak fields
might not be unrealistic within the framework of cold atoms in optical lattices or ion traps 
(see, e.g. \cite{Bloch,Bogdan-Rev}).

\begin{figure}[h]
\centerline{
\includegraphics[width=0.8\linewidth, angle=0]{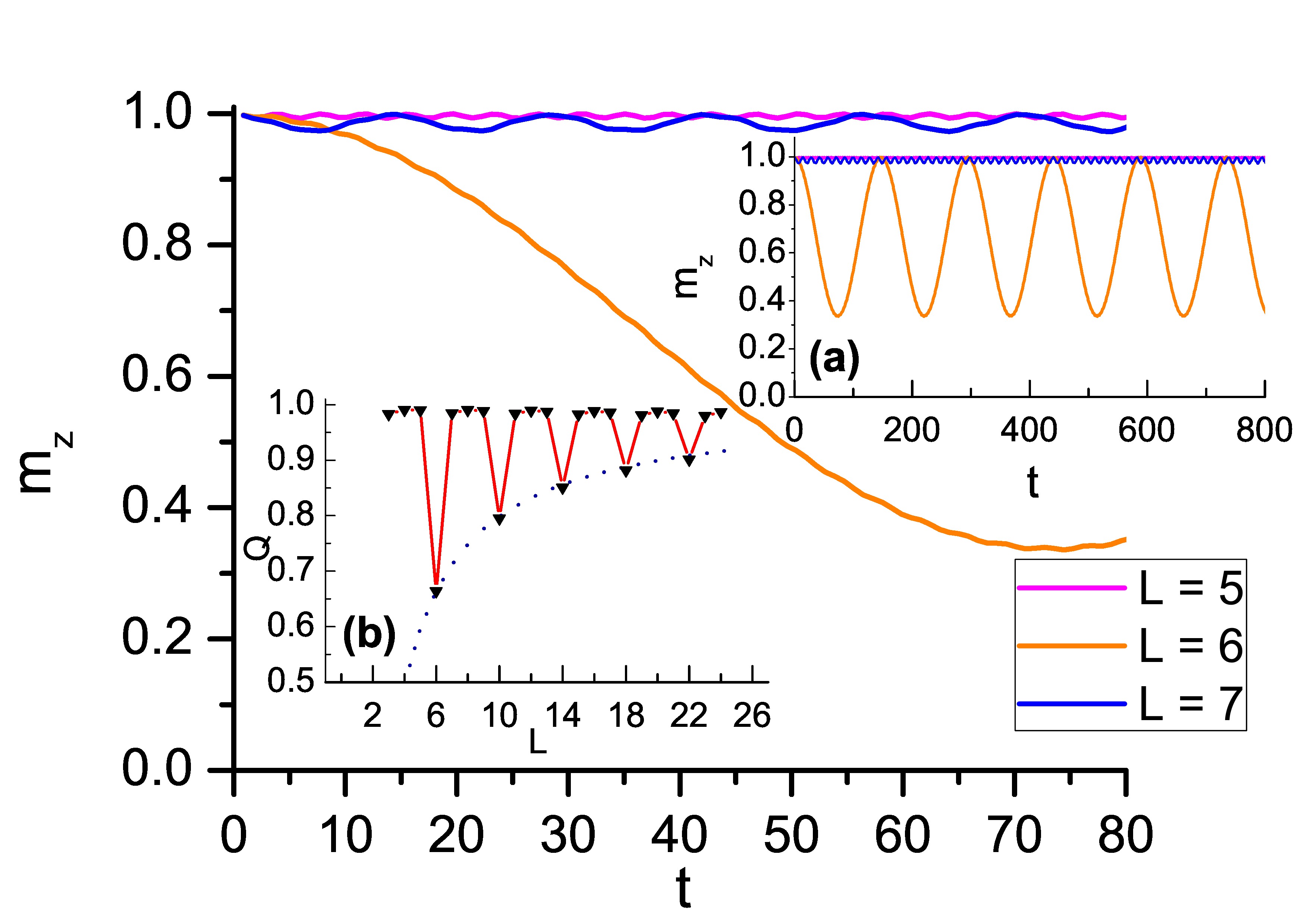}
}
\caption{\small
{\bf{Manifestation of dynamical transition at small system-size ($L$):}} 
The figure (main + inset {\bf (a)}) shows (numerical results) drastic contrast of response dynamics 
for $L = 5$ and $7$ with that for $L = 6$ for $\delta = \gamma = 1$, $\omega = \omega^P_{0} + 10^{-2}$.
Even for $\delta \ne 0$, for small $L$ strong dynamics is observed 
for certain system-sizes ($L = 2(2n+1)$, $n = 1,2,3, ...$) represented by the corresponding dips in $Q$
in inset {\bf (b)}. These are the system sizes for which a mode $k$ has $\cos{k} = 0$, 
which will have full oscillation independent of the value of $\omega$ (Eqs. \ref{Vk2}, \ref{phik}). 
The blue dotted line is $Q = 1 - r/L$ with $r$ is determined using the numerical value of $Q$ for $L=6$ (the first dip). 
}
\label{Q-vs-L}
\end{figure}
\noindent
{\it{\bf Experimental possibilities: finite-size versions}}
Various kinds of quantum spin Hamiltonians and their 
coherent dynamics are realizable in cold atomic systems
in optical lattice (see. e.g., \cite{Bloch,Bogdan-Rev}). One important
feature of the switching phenomenon that makes it particularly convenient for
experimental realization is its clear signature
even at very small system sizes of just a few components  
(with periodic boundary condition). 
Suppose we are away from the transition condition (i.e. $\delta \ne 0$), and drive 
the system with  $\omega,h_{0} \gg \gamma, \delta$ close to the large system perfect 
freezing condition: some small systems,
$L = 3, 4, 5, 7, 8, 9$ also freeze. However, at $L = 6$ we see pronounced dynamics in the 
system, resulting in a strong dip in $Q$. Similar pronounced dynamics are observed at system sizes 10, 14 ... $2(2n+1)$ with gradually 
decreasing amplitude (Fig. \ref{Q-vs-L}). 

The origin of such strong dynamics even for $\delta \neq 0$ can be gleaned from Eqs. (\ref{Vk2}) and (\ref{phik}).
For $L = 2(2n+1)$ we have a pair of $k$-modes for which $\delta \cos{k} = 0$: again the term involving the Bessel 
factor responsible for the freezing cancels out for that particular pair of modes
and they execute full oscillation. For small $L$, there are only few modes in total, and hence
the contribution from the fully oscillating modes is substantial resulting in the dips in $Q$. 
As $L$ increases, $Q$ at the dips scales as $Q(L) = 1 - r/L$, where
$r$ is a constant (inset of Fig. \ref{Q-vs-L}). 

Since this small system-size signature of the transition 
does not require $\delta \rightarrow 0$, it can also be observed even in the 
transverse Ising chain, which is a special of the 
XY chain with $\delta = \gamma = 1$. Indeed, shown in the figure are results 
for the Ising case where in fact no dynamics
is actually expected in the large $L$ limit, since in this case we have $\gamma=\delta = 1$.
We emphasize that the transverse Ising chain with time-varying 
transverse field has already been realized 
experimentally \cite{Kim, Schatz}.
\noindent
{\bf Different initial conditions:} 
Freezing at $\omega^P$ is maximal as there,
every mode freezes 
individually \cite{AD-DQH}, 
while for $\delta=0$, every mode becomes dynamical. In that sense, freezing and 
switching are independent of the starting state, which only affects  
the amplitudes $u_{k}(t=0)$ and $v_{k}(t=0)$, which can be accounted for 
by a time-independent unitary transformation. Such a transformation preserves 
$Q = 0$, although full population transfer would not occur in the
modified basis. Thus switching and the concomitant discontinuous limit are essentially
unaffected but the fully polarized initial state we have chosen
provides a large $Q$ away from the transition point, making the transition
most clearly visible.

\noindent
{\bf Conclusion and outlook:} We have  presented switching between (completely) frozen and dynamical 
non-equilibrium regimes under driving, 
with regime change effected by simply tuning a parameter in the Hamiltonian. 
This set of phenomena is particularly intriguing on account of how fast driving 
yields the appearance of long timescales non-monotonically. Given the applicability 
of the basic Hamiltonian studied here to a wide range of systems, and the appearance of switching even 
for small systems, experimental realisation does not appear an unrealistic prospect in the near future. 

Several further questions obviously spring to mind. Firstly, what are the necessary conditions 
for such behaviour to occur beyond the soluble RWA -- here, e.g., the point $\delta=0$ is special 
in that it exhibits an enhanced XY-like symmetry, as well as a degenerate Floquet spectrum, and 
appears like a critical point with a diverging timescale -- but it is not clear which of these
fundamentally underpin such switching mechanisms in general.  
As a next step, 
progress towards studying the long-time dynamics of non-integrable systems would be of interest to 
determine the stability of this phenomenology, this being hampered by the present inability to 
access the long-time dynamics of such systems. Finally, given the growing zoo of such nonequilibrium 
quantum phenomena, a systematic zoology based on some yet to be identified organising principles  
is clearly a most desirable goal in the long term. 



\begin{thebibliography}{99}
\bibitem{Non-Symmetric}
If the equilibrium (quasi-static) average of the response
over a driving period is non-zero (say due to some equilibrium background
value), one subtracts that quasi-static average from the 
response while defining the anomalous part of the dc response.
\end{thebibliography}
\end{document}